\documentclass[thmsa,onecolumn,12pt]{article}%

\usepackage{authblk}
\usepackage{graphicx}
\usepackage{multirow}
\usepackage{subfigure}
\usepackage{amsfonts}
\usepackage[a4paper,left=2.4cm,right=2.3cm,top=2cm,bottom=1.8cm]{geometry}%
\usepackage{amsmath,amssymb,mathrsfs}
\usepackage{amsthm}
\usepackage{color}

\usepackage{hyperref}
\usepackage{float}
\usepackage{epsfig}
\usepackage{epstopdf}
\usepackage{array,booktabs}
\usepackage{tabularx}
\usepackage[pagewise]{lineno} 

\providecommand{\U}[1]{\protect\rule{.1in}{.1in}}

\makeatletter

\newcommand{\Rmnum}[1]{\expandafter\@slowromancap\romannumeral #1@}
\makeatother

\newtheorem{remark}{Remark}

\begin{document}

\title{Estimating Adsorption Isotherm Parameters in Chromatography via A Virtual Injection Promoting Feed-forward Neural Network}

\author[1]{Chen Xu}
\author[2,1]{Ye Zhang}
\affil[1]{Shenzhen MSU-BIT University, 518172 Shenzhen, China}
\affil[2]{School of Mathematics and Statistics, Beijing Institute of Technology, 100081 Beijing, China}
\date{}

\maketitle

\begin{abstract}
The means to obtain the adsorption isotherms is a fundamental open problem in competitive chromatography. A modern technique of estimating adsorption isotherms is to solve an inverse problem so that the simulated batch separation coincides with actual experimental results. However, this identification process is usually ill-posed in the sense that the small noise in the measured response can lead to a large fluctuation in the estimated quantity of adsorption isotherms. The conventional mathematical method of solving this problem is the variational regularization, which is formulated as a non-convex minimization problem with a regularized objective functional. However, in this method, the choice of regularization parameter and the design of a convergent solution algorithm are quite difficult in practice. Moreover, due to the restricted number of injection profiles in experiments, the types of measured data are extremely limited, which may lead to a biased estimation. In order to overcome these difficulties, in this paper, we develop a new inversion method -- the Virtual Injection Promoting Feed-forward Neural Network (VIP-FNN). In this approach, the training data contain various types of artificial injections and synthetic noisy measurement at outlet, generated by a conventional physics model~-- a time-dependent convection-diffusion system. Numerical experiments with both artificial and real data from laboratory experiments show that the proposed VIP-FNN is an efficient and robust algorithm.
\end{abstract}

%

\section{Introduction}
\label{sec:Introduction}

Chromatography is a useful approach in separation/purification processes to isolate one or several components from a mixture. This technique is particularly important when the mixture components are difficult to separate by traditional processes, such as distillation or extraction, both in industry and in academia~\cite{GuiochonLin-03}. Its mechanism is based on the fact that different solutes in the sample interact differently comparing with the stationary phase. The outcome of a chromatographic separation is strongly dependent on the adsorption isotherms of the solutes, because they dictate the separation factors and saturation capacities. Thus, how to obtain adsorption isotherms is an important question in chromatography~\cite{ForssenArnell-06}.

There are several experimental methods that can be used to measure adsorption isotherms, e.g. the perturbation peak~\cite{DoseJacobson-91}, the frontal analysis~\cite{Lisec-01}, etc. Another line of research, based on solving an inverse problem through the computer simulations, have recently gained much attention, see e.g. ~\cite{FelingerZhou-03,ForssENFornstedt-15,ForssenArnell-06,zhang2016regularization}. This type of method is used to numerically estimate adsorption isotherm parameters so that the simulated batch separation coincides with actual experimental results. Since this method only requires a few injections of different sample concentrations, solute consumption and time requirements are very modest. However, from a mathematical viewpoint, such an inverse problem, i.e. estimating adsorption isotherm from measured data, is a typical ill-posed problem in the sense of Hadamard. Therefore, regularization methods should be employed for obtaining meaningful approximate solutions. For this inverse chromatography problem, Tikhonov-type variational regularization approaches are usually used in the literature, see e.g. ~\cite{ChengLin2018,James-94,zhang2016regularization,LinZhang2018}. However, there are some drawbacks for such methods. One is that due to the nonlinearity of inverse chromatography problem, the formulated optimization problem is highly non-convex. Therefore, the question of how to design a convergent solution algorithm for the corresponding non-convex optimization problem is an open problem. Gradient decent algorithms are the most important candidates for such assignment. However, as shown in \cite{zhang2016}, the calculation of gradients of the objective functional involves solving an adjoint problem, which is unstable and computationally expensive. Another difficulty is that the prior information needs to be encoded as an explicit functional, which limits the type of a priori information that can be accounted for. Moreover, how to appropriately choose the regularisation parameter is non-obvious for our inverse chromatography problem in practice. Finally, due to the limitation of injection pattern in the measurement data, the conventional least square based approaches are usually biased and strongly dependent on the injection profile in the experiments.  In this work, in order to avoid these difficulties we propose a machine learning approach to estimate the adsorption isotherms by using the artificial full injection profiles.

Machine learning, a data-driven approach, has been increasingly applied to many research fields in the past decades, such as finance and accounting(e.g. \cite{Gu-Kelly-Xiu-2020, Freyberger2020, Zhou2011}),  computer vision (e.g. \cite{He-Zhang2016,Shin2016}), and machine translation (\cite{Vaswani2017, devlin2019bert}). Its impact on the field of inverse problems is also growing (\cite{DeVito2005, Lucas2018, Raissi2018, Morshed1998}). Machine learning solves the problem of optimizing a performance criterion based on statistical
analyses using example data or past experiences \cite{Hastie2009}. In particular, methods that combine forward modelling with data driven techniques are being developed~\cite{Arridge-Maass-Oktem-Schonlieb}. Some of these techniques build upon the similarity between deep neural networks and classical approaches to inverse problems such as iterative regularization~\cite{Adler-Oktem-2017}. Some are based on postprocessing of the reconstructions obtained by a simple inversion technique such as filtered backprojection~\cite{Jin-McCann-2017}. Others use data driven regularizers in the context of variational regularization~\cite{Li-Schwab-2020} or use deep learning to learn a component of the solution in the null space of the forward operator~\cite{Schwab-Antholzer-2019,Bubba-Kutyniok-2019}. Recently, in the field of numerical Partial Differential Equations (PDEs), many machine learning algorithms has been proposed to solve forward and inverse problems for PDEs, see e.g.~\cite{HanJentzenE2018,LiChenTaiE2018,HeXu2019,HeLiXuZheng2020}. Specifically, a physics-informed neural network (PINN) was proposed in \cite{Raissi2019} for solving inverse problems in PDEs. Its essential idea is to infer the unknown solution (physics of interest) by combining the governing equation and the given data (e.g., initial/boundary conditions or partial and scattered measurements of the any of the states). In the present work, we aim to adopt the idea of PINN to learn the states of interest, i.e. the adsorption isotherms.

Feed-forward Neural Network (FNN) is one of the most-widely used models in machine learning. Similar to the well-known linear regression models, it can be used to approximate relationships between variables. The difference is that besides the linear ones, FNN is also good at describing nonlinear relationships. Theoretical works (see, e.g. \cite{Kurt1990}) show that under mild conditions, the neural network is able to approximate an arbitrary continuous function. This result has been verified by a large amount of empirical evidences where neural networks are capable in describing complex relationships (\cite{Gu-Kelly-Xiu-2020, alfaro2008, fletcher1993, Morshed1998}). Based on these evidences and the idea of PINN, in this paper, by exploiting the potential injection information, we develop a specific FNN, named as Virtual Injection Promoting Feed-forward Neural Network (VIP-FNN), for estimating adsorption isotherm parameters. Another motivation to choose FNN over other candidate models, e.g. linear models, decision trees (RT), and support vector machines (SVM)\footnote{See \cite{Hastie2009} for a thorough discussion of the tree model and SVM.}, is that it is good at describing complex relationships\footnote{We expect the relation between independent and dependent variables in our problem to be complex since there does not exist an analytic function between them.} and is tolerant to noises\footnote{Our final model will be applied in lab experiments where measurement errors are expected.}.

The remainder of this paper is structured as follows. Section 2 aims to give  the  background of forward and inverse problems in liquid chromatography and some classical approaches for solving the corresponding inverse problem. In Section 3, we reformulate the corresponding inverse problems in the language of data science, and solve it by the developed machine learning approach -- VIP-FNN. Computer simulations for both synthetic problems and real-world problems are demonstrated in Section 4. Finally, concluding remarks are given in Section 5.

\section{Conventional Mathematical Models for Forward and Inverse Problems in Chromatography}
\label{basis}

In this section we briefly review the conventional mathematical models in competitive liquid chromatography related to adsorption isotherm estimation problem. First, let us recall the commonly used chromatographic model in a fixed bed chromatography column. To this end, denote by $C$ and $q$ concentrations in the mobile and the stationary phase, respectively. Then, if the mass transfer kinetics and column efficiency are sufficiently high, the migration of the molecules through a chromatographic column can be modeled by the  following time-dependent convection-diffusion system (refer to \cite{LinZhang2018} for a simple derivation)
\begin{equation}\label{eq:prime}
\left\{\begin{array}{ll}
\frac{\partial C}{\partial t}+F\frac{\partial q}{\partial t}+u\frac{\partial C}{\partial x}=D_{a}\frac{\partial^{2} C}{\partial x^{2}}, & x\in (0,L],~ t\in (0,T], \\
C(x,0)=g(x),& x\in (0,L], t=0, \\
C(0,t)-\frac{D_{a}}{u}\frac{\partial C(0,t)}{\partial x}=h(t), & x=0, t\in (0,T],\\
D_{a}\frac{\partial C(L,t)}{\partial x}=0,& x=L, t\in (0,T].
\end{array}\right.
\end{equation}
where $u$ denotes the mobile phase velocity, $F$ represents stationary/mobile phase ratio, and $D_{a}$ is the diffusion parameter. In this work, all of $u, F$ and $D_{a}$ can be assumed as fixed numbers. $L$ is the length of chromatographic column, and $T$ is a appropriate time point slightly larger than the dead time of chromatographic time $T_0 = L/u$. Further, $x$ is distance, $t$ is time, $g$ is the initial condition and $h$ is the boundary condition, which describes the injection profile of the problem. For the problem with $n$ components, $C$, $q$, $g$ and $h$ are vector functions of size $n$, where $n$ denotes the number of components. For the competitive chromatography, $n\geq2$.

When the mass transfer resistance for adsorption/desorption is small, i.e. the fast kinetics, the quantity $q=q(C)$ in \eqref{eq:prime}, termed as the adsorption isotherm, describes the relationship between the amount of the component in the mobile phase and the amount of the component adsorbed on the stationary phase. In the multi-component preparative situation, $q(C)$ is a nonlinear function of $C$ since each component is a function of all component concentrations because of competition for access to the adsorption sites. There are several methods of mathematically representing adsorption isotherms $q$, with different models used to describe the adsorption process~\cite{Guiochon-06,GuiochonLin-03}. In this work, we consider the following commonly used Bi-Langmuir isotherm model [14]
\begin{equation}\label{BiLangmuir}
q^\ast_\mu(C;\mathbf{y})=\frac{a_{\Rmnum{1},\mu}C_\mu}{1+\sum^{n}_{j=1}b_{\Rmnum{1},j}C_{j}} + \frac{a_{\Rmnum{2},\mu}C_\mu}{1+\sum^{n}_{j=1}b_{\Rmnum{2},j}C_{j}}, \quad \mu=1,\cdots,n,
\end{equation}
where the vector $\mathbf{y}\in \mathbb{R}^{8}$ denotes as a collection of all parameters $a_{\nu,\mu}$ and $b_{\nu,\mu}$, describing the adsorption of the component $\mu$ to the site $\nu$ ($\nu=\Rmnum{1},\Rmnum{2}$). In the model \eqref{BiLangmuir}, the first of the two Langmuir components can be identified as accounting for the nonselective interactions and the second for the selective interactions between the enantiomers and the enantioselective adsorbent. To be more precise, the $a$ terms are related to the chromatographic retention factor and dictate adsorption under linear conditions, while the $b$ terms are the thermodynamic association constants for the respective binding sites.

Traditional methods of determining the parameter $\mathbf{y}$ (see e.g.~\cite{zhang2016regularization,zhang2016} and references therein) is to find a solution $C(x,t;\mathbf{y})$ of PDE (\ref{eq:prime}) so that its value at the outlet $C(L,t;\mathbf{y})$ coincides with the actual experimental results $C_{obs}(t)$. The mathematical formulation of the above identifying process reads the following least square problem
\begin{equation}\label{optimization0}
\min_{\mathbf{y}\in \Theta}  \frac{1}{2} \left\| C(L,t;\mathbf{y})- C_{obs}(t) \right\|^2_{(L^2[0,T])^n},
\end{equation}
where $C(\cdot,t;\mathbf{y})$ solves PDE \eqref{eq:prime} with a given parameter $\mathbf{y}$ and $C_{obs}(t)=(C_{1,obs}(t), ..., C_{n,obs}(t))$ are observed concentrations for all components at the column outlet at time t. The set $\Theta$ contains the \emph{a priori} information of $\mathbf{y}$, e.g. the non-negativity constraint $\mathbf{y}\geq0$.

Note that in almost all cases what is observed at the column outlet is not concentrations of each component $\{C_{\mu,obs}(t)\}^n_{\mu=1}$ but instead, the total response $\mathbf{r}^{obs}\in \mathbb{R}^{N_T}$ at time grid $\{t_i\}^{N_T}_{i=1}$ is measured, where we have that,
\begin{equation*}
\mathbf{r}^{obs}_i = \min \left\{ \sum^n_{\mu=1} f_{cal,\mu} (C_{\mu,obs}(t_i)) , r_{max} \right\}, \quad i=1, \dots, N_T,
\end{equation*}
where $f_{cal,\mu}$ is a calibration function giving components' detector response as a function of component concentration and $r_{max}$ is the detector's saturation limit. Denote
\begin{equation}
r_{tot} (C(L,t_i;\mathbf{y})) = \min \left\{ \sum^n_{\mu=1} f_{cal,\mu} (C_{\mu}(L,t_i;\mathbf{y})) , r_{max} \right\}
\label{TotalResponse}
\end{equation}
as the total response of simulated concentration at the outlet. Then, the least square problem (\ref{optimization0}) becomes
\begin{equation}\label{optimization2}
\min_{\mathbf{y}\in \Theta} \sum^{N_T}_{i=1} \left( r_{tot}(C(L,t_i;\mathbf{y}) )- \mathbf{r}^{obs}_i \right)^2.
\end{equation}

The formulation \eqref{optimization2} is still ill-posed since the uniqueness of the minimizer
cannot be guaranteed by the integral measured data  $\mathbf{x}$ (note that the objective functional \eqref{optimization2} remains constant if one moves $C(L,t;\mathbf{y})$ along the $t$-axis when $supp(r_{tot}(C(L,\cdot;\mathbf{y}))\cap r_{obs}(\cdot)) = \emptyset$~\cite{James-94}). For overcoming the ill-posedness, the first momentum regularizing strategy is usually adopt in the solution method, i.e. the approximate adsorption isotherm parameters are designed as the minimizer of the following optimization problem
\begin{equation}\label{optimization3}
\min_{\mathbf{y}\in \Theta} \sum^{N_T}_{i=1} \left( r_{tot}(C(L,t_i;\mathbf{y}) )- \mathbf{r}^{obs}_i \right)^2 + \alpha  \left\{ \sum^{N_T}_{i=1} t_i\left( r_{tot}(C(L,t_i;\mathbf{y}) )- \mathbf{r}^{obs}_i \right) \right\}^2,
\end{equation}
where $\alpha>0$ is the regularization parameter.

It should be noted that the above modeling is based on one trial with a fixed injection profile. In practice, in order to make the inversion model more reliable, several experiments with different  injections are proceeded. In this case, the inject enhanced optimization model becomes
\begin{equation}\label{optimizationInjections}
\min_{\mathbf{y}\in \Theta} \sum^{N_\sigma}_{\sigma=1} w_\sigma \sum^{N_T}_{i=1} \left( r_{tot}(C^\sigma(L,t_i;\mathbf{y}) )- \mathbf{r}^{obs,\sigma}_i \right)^2 + \alpha \sum^{N_\sigma}_{\sigma=1} w_\sigma  \left\{ \sum^{N_T}_{i=1} t_i\left( r_{tot}(C^\sigma(L,t_i;\mathbf{y}) )- \mathbf{r}^{obs,\sigma}_i \right) \right\}^2,
\end{equation}
where $C^\sigma(L,t;\beta)$ and $\mathbf{r}^{obs,\sigma}$ are simulated concentration and observation data, corresponding to different trails with the injection function $h^\sigma(t)$. Coefficient $w_\sigma$ in (\ref{optimizationInjections}) normalizes the experimental elution profiles. If the elution profiles are not normalized, the optimization routine will be biased towards determining adsorption isotherm parameters that provide a good fit to high concentration elution profiles. Usually, the weights $w_\sigma$ are chosen, so that
\begin{equation*}\label{normalized}
w_1  \sum^{N_T}_{i=1} \left( \mathbf{r}^{obs,1}_i \right)^2 = \cdot\cdot\cdot = w_{N^\sigma}  \sum^{N_T}_{i=1} \left(  \mathbf{r}^{obs,N_\sigma}_i \right)^2.
\end{equation*}

In (\ref{optimizationInjections}), $N_\sigma$ presents the number of trails (it equals the number of testing injection functions). Clearly, more additional information about the physics problem will help us to obtain a better result, i.e. the bigger the $N_\sigma$, the more trustable of the estimated adsorption isotherm parameters. Actually, this is the main drawback of the conventional solution method, cf. (\ref{optimizationInjections}), which strongly depends on the injection profiles of real data. Other shortcomings of the conventional solution method are (a) the regularization parameter $\alpha$ is difficult to choose in practice, and (b) there is no efficient algorithm to find an even local minimizer of  \eqref{optimizationInjections}. In order to overcome these three drawbacks, we will propose a powerful machine leaning method in next section, which uses the artificial experiments to raise the injection information for estimating the adsorption isotherm parameters.

\section{Virtual Injection Promoting Feed-forward Neural Network (VIP-FNN) for Estimating Adsorption \\ Isotherm Parameters}
\label{FNN}

The goal in this section is to develop an injection informed machine learning method -- VIP-FNN -- for estimating adsorption isotherm parameters in competitive liquid chromatography. Without loss of generality, we consider in this paper the chromatographic system with two components, i.e. $n=2$.

Denote the injection informed data $\mathbf{x}\in \mathbb{R}^{N_T+2}$ of a chromatographic system as
\begin{equation}
\mathbf{x} := (\mathbf{r}^{obs}, \bar{h}_1, \bar{h}_2),
\label{y}
\end{equation}
where $N_T$ denotes the size of $\mathbf{r}^{obs}$ and $\bar{h}_\mu$ denotes the injection profile of component $\mu(=1,2)$. The dynamical injection function $h(t)=[h_1(t); h_2(t)]$ has the form $h_\mu(t)= H(10-t) \bar{h}_\mu$, where $H(\cdot)$ denotes the Heaviside step function. The main advantage of VIP-FNN over traditional inversion methods, cf. \eqref{optimizationInjections}, is that during its construction it uses all possible injection profile $\bar{h}_\mu$ (though it is a physics model derived with an artificial pattern), which is sparse partially-available information if we only consider the real experimental data.

Now, we are in the position to present the steps in our algorithm to develop VIP-FNN. We skip the construction procedure for FNN as it is a standard method in machine learning. Alternatively, a brief introduction to FNN is given in the appendix, for more details refer to \cite{Bishop06,Svozil1990}.

\subsection{Building the Pretrained-Model on the Error-free Data}
\subsubsection{Data Structure}
\label{data_structure}
Let $X\in \mathbb{R}^{N_T+2}$ be the vector of independent variables, where the first $N_T$ variables represent the measurement data at the outlet at the time grid $\{t_i\}^{N_T}_{i=1}$ and the last two variables are the two injection components. In this work, we set $N_T=800$. Let $Y\in \mathbb{R}^{8}$ be the vector of dependent variables, representing the eight adsorption isotherm parameters. Assume that both $X$ and $Y$ are random vectors. The problem of recovering adsorption isotherm parameters from measurement data is transferred to problem of estimating the function from $X$ to $Y$, which will be solved by FNN in this work.

Each entry of $X$, i.e. $X_i, i=1, \cdots, N_T+2$ will be called a \emph{feature}. For $k=1,2,...,N$, let $\mathbf{x}_k=(x_{1,k}, \cdots, x_{N_T+2,k})^T$ be the $k$-th realization of $X$. It is a $N_T+2$ by 1 deterministic vector. Similarly, let $\mathbf{y}_k = (y_{1,k}, \cdots, y_{8,k})$ be the $k$-th realization $Y$ corresponding to $\mathbf{x}_k$.

The samples, $\{(\mathbf{x}_k,\mathbf{y}_k)\}_{k=1}^N$ where $N = 63500$ in our work, are generated in the following way (according to the experience of experts in the field):
\begin{itemize}
\item For each of $j=1,2,...,8$, $\{y_{j,k}\}_{k=1}^N$ are sampled independently from Uniform(0,100);
\item For $i=801$ and $802$, $\{x_{i,k}\}_{k=1}^N$ are sampled independently from Uniform(0,30);
\item For each sample $k$, the last two entries of $\mathbf{x}_k$, $x_{801,k}$ and $x_{802,k}$, are already obtained in the last step. The first 800 entries of $\mathbf{x}_k$, $\{x_{i,k}\}_{i=1}^{800}$, are computed by numerically solving PDE \eqref{eq:prime} with the given parameters $x_{801,k}$, $x_{802,k}$ and $\mathbf{y}_k$ (the PDE \eqref{eq:prime} is solved by finite volume method, presented in \cite[\S 6.1.1]{zhang2016regularization}). It should be noted that for measurements with different time grids, we apply piecewise Hermite polynomials interpolation to obtain the modified measurements on the same time grid.
\end{itemize}

\begin{figure}[!htb]
	\centering
\includegraphics[scale=.43]{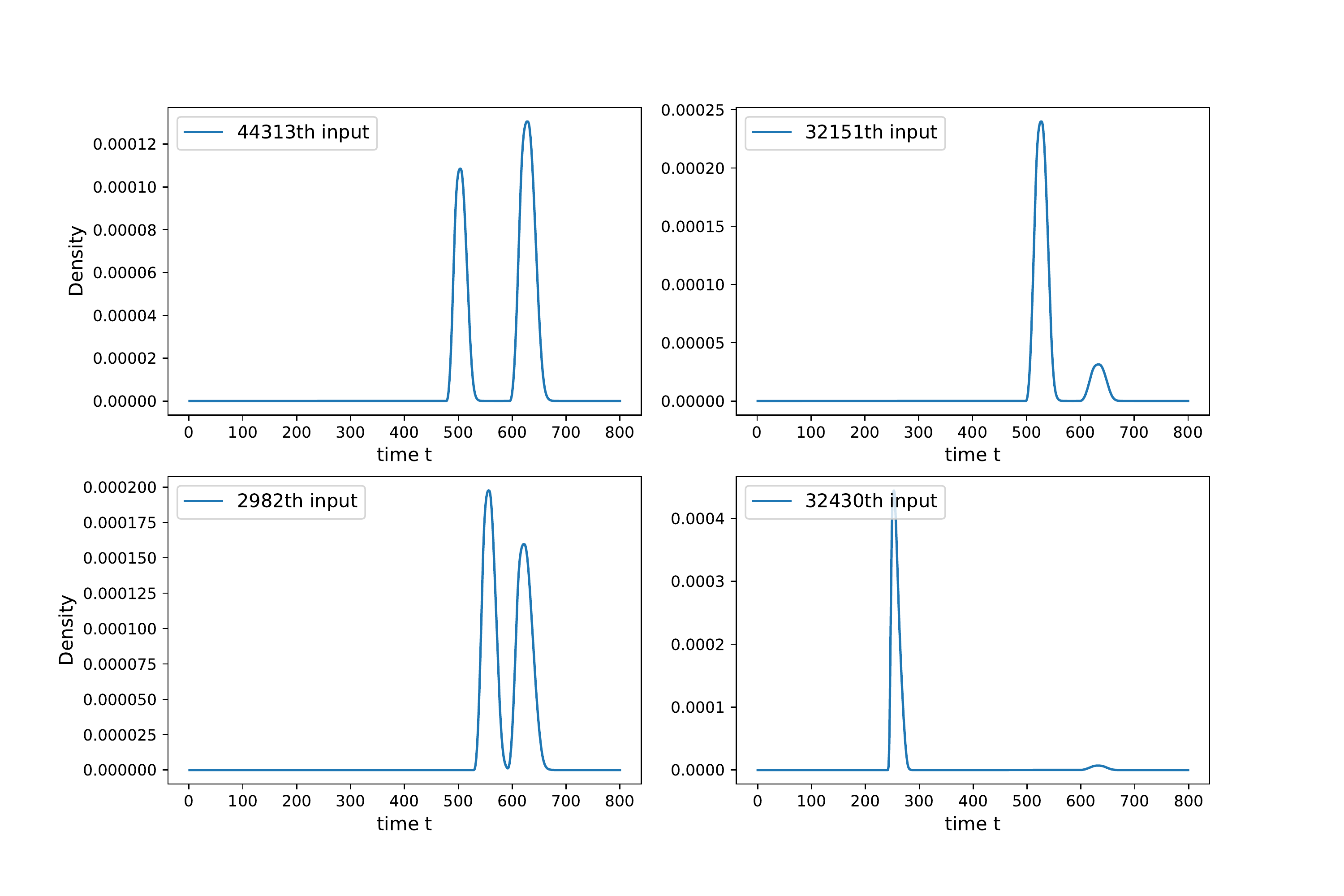}
\caption{Error-free data: we randomly choose four samples and plot their inputs, each of which is a time series of densities.}
\label{inputplot}
\end{figure}

\begin{remark}
We call the above as error-free data, since we will add noise (errors) to them later in Section \ref{noisedata}. To get a sense on what these error-free data look like, we plot the inputs (excluding the last two entries) of four randomly selected samples in Figure \ref{inputplot}.
\end{remark}

It is a standard procedure to scale the inputs before feeding them to the model for training or testing. Some of the main reasons are, it makes the model easier to train and leads to a higher likelihood of finding the optimal weights. Hence, for each $i=1,2,...,802$, we normalize the training samples of $X_i$ so that they have zero-mean and a standard deviation of one. Specifically, we do the following before feeding data to the model.

\begin{itemize}
\item For each $i=1,2,...,802$, we calculate the sample mean and standard deviation for the training samples:
\[\bar{x}_i:=\sum_{k=1}^{N_{tr}}x_{i,k}/{N_{tr}}, \quad \sigma(x_i):= \sqrt{\sum_{k=1}^{N_{tr}}(x_{i,k}-\bar{x}_i)^2/{N_{tr}}},\]
where $N_{tr}=38100$ is the training sample size which we set as $60\%$ of the total samples.
\item For each $x_{i,k}$ in all the training, we replace it with $(x_{i,k}-\bar{x}_i)/\sigma(x_i)$. Note that before using the trained model to do predictions on any new sample (for example, the validation or the testing samples)  input $\{z_i\}_{i=1}^{800}$, we also replace $z_i$ with $(z-\bar{x}_i)/\sigma(x_i)$.
\end{itemize}

\begin{figure}[!b]
	\centering
		\includegraphics[scale=0.7]{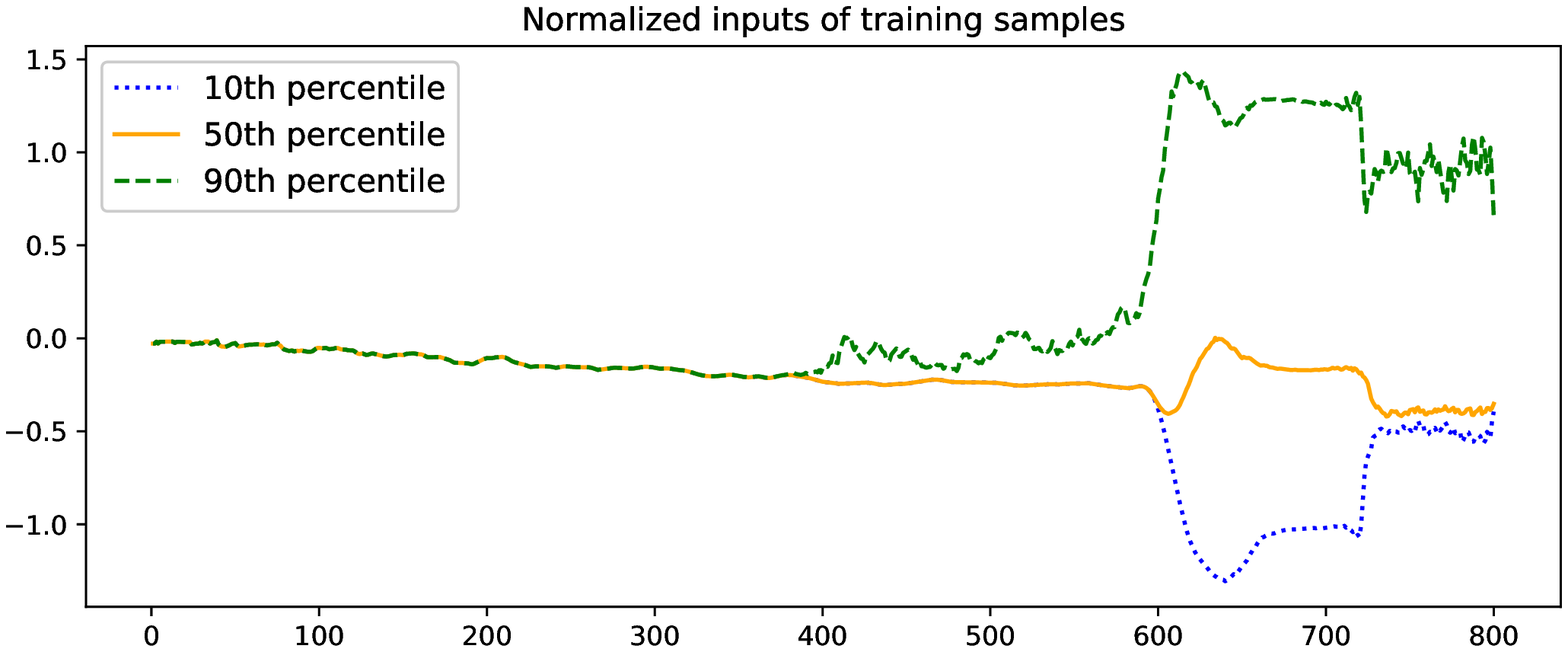}
		\includegraphics[scale=0.7]{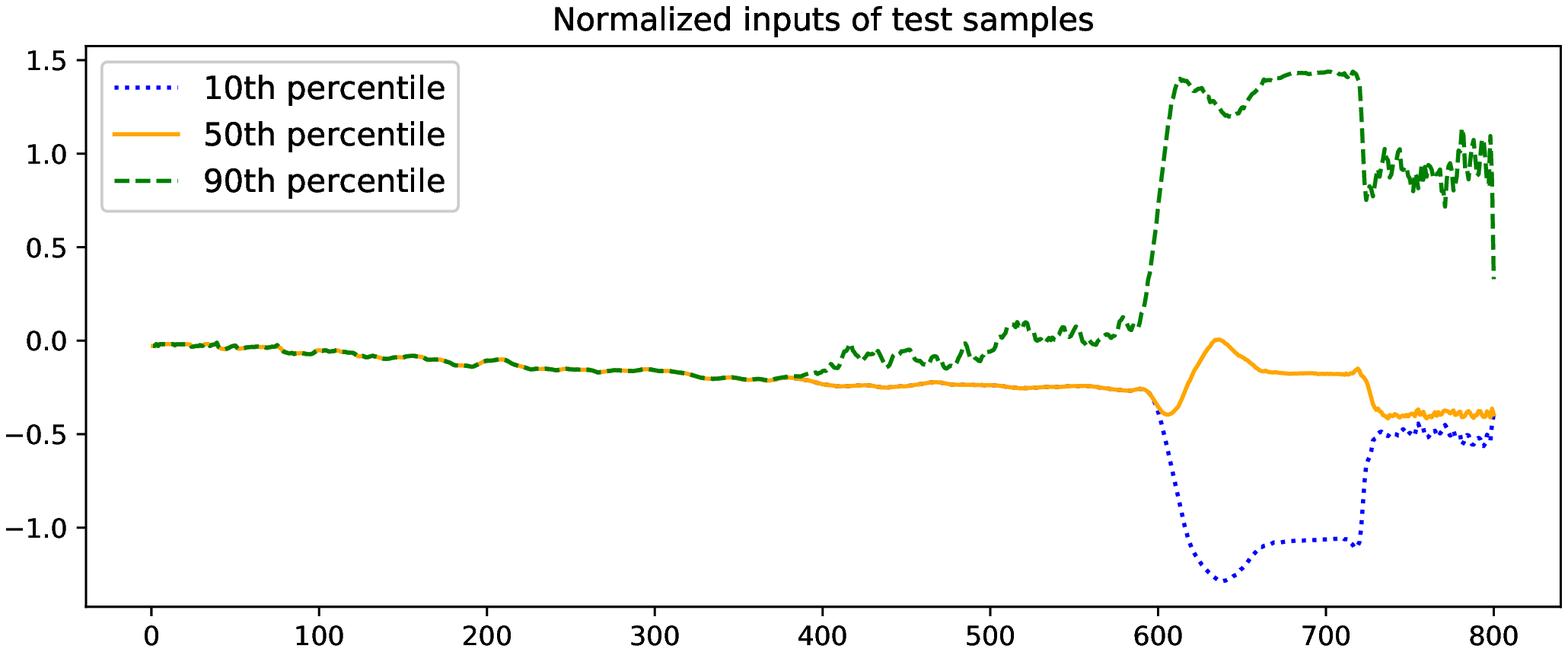}

\caption{Percentiles of the normalized inputs}
\label{normalized-inputs}
\end{figure}

For $i=1,2,...,800$, the 10th, 50th, and 90th percentiles of the normalized inputs, $\{x_{i,k} \}_{k=1}^{N_{tr}}$, in the training samples are plotted in Figure \ref{normalized-inputs}. It also includes the percentiles for inputs in the testing samples. The plots for the training and testing samples are almost of the same shape. Note that the last two entries of the normalized inputs, i.e. the two injection components, are not included in the plot because they are data of different kinds from the first 800 entries.

\subsubsection{Loss, Metric, and Activations of the Neural Net}
For each $\mathbf{x}_k$ ($k=1,\cdots,N$), denote the output of the neural net by $\hat{\mathbf{y}}_k$. Our loss function consists of the estimation error and regularization terms for weights $\{w_j\}$ and bias terms $\{b_p\}$  (see Figure \ref{FNNexample} in the appendix for illustration of the bias terms). There are two common choices on the norm function for the error term and the regularization terms: the mean-absolute-error (MAE, $L1$):
\begin{equation}
L1 \text{ Loss}: \mathfrak{L} = \frac{1}{N'}\sum_{k=1}^{N'} |\hat{\mathbf{y}}_k-\mathbf{y}_k| + \alpha_w\sum_{j=1}^J |w_j| + \alpha_b \sum_{p=1}^P |b_p|,
\label{L1}
\end{equation}
and the mean-square-error (MSE, $L2$):
\begin{equation}
L2 \text{ Loss}: \mathfrak{L} = \frac{1}{N'}\sum_{k=1}^{N'} |\hat{\mathbf{y}}_k-\mathbf{y}_k|^{2} + \alpha_w\sum_{j=1}^J w_j^2 + \alpha_b \sum_{p=1}^P b_p^2.
\label{L2}
\end{equation}
We will choose the one that performs better on the training and validation datasets. In \eqref{L1} and \eqref{L2}, $\alpha_w$ and $\alpha_b$ are hyper-parameters, which are also called the regularization double-parameters in the community of inverse problems, $N'(\leq N)$ is the size of the sample set under consideration, $w_j$'s are the weights, $J$ is the total number of weights, $b_p$'s are the bias terms, and $P$ is the total number of biases. $|\cdot|^p$ denotes the mean-$p$-norm, i.e.
\[ \forall \mathbf{y}=[y_1; \cdots; y_M]\in \mathbb{R}^M: ~ |\mathbf{y}|^p := \frac{1}{M}\sum_{i=1}^{M} |y_i|^p.\]
The goal for the later model training process is to find proper values of $\{w_j\}$ and $\{b_p\}$ to minimize the loss $\mathfrak{L}$. The reason to include the regularization term in the loss is to restrict the size of the weights (also biases) and hence to prevent overfitting problems\footnote{Overfitting refers to the case when the model captures features that only belong to the training data, which reduces the model's capacity on new data. See \cite[\S 11.5.2]{Hastie2009} for more details.}.

We choose the $R^2$ statistic, a standard metric for regression, to evaluate our model's performance on any sample with a size $N'$:
\[ R^2 := 1 - \frac{\sum_{k=1}^{N'} |\hat{\mathbf{y}}_k-\mathbf{y}_k|^2}{\sum_{k'=1}^{N'}|\bar{\mathbf{y}}-\mathbf{y}_{k'}|^2}, \]
where $\bar{\mathbf{y}}:=\sum_{k=1}^{N'}\mathbf{y}_k/N'$.
The possible range for $R^2$ is $(-\infty,1]$ and the closer $R^2$ gets to 1 the better the model is. Note that although the loss $\mathfrak{L}$ is close-related to the model's prediction capacity, it is not a proper measure of performance, since it involves regularization terms which do not reflect the model performance.

Finally, in our FNN, the activation function will be selected by trials from among the widely-used tanh and sigmoid, based on the mean-square-error values on the validation data.

\begin{remark}
\label{remarkReLu}
Note that we do not include the rectified linear unit (ReLU) in our pool of activations due to its ``dead-neuron'' problem, i.e. during the training, when the ReLU activation function is used, if a neuron is not activated in some step, it will never be activated in all the following steps even if it should be activated in the true model.
\end{remark}

\subsubsection{Building the Neural Net}
\label{buildmodel}
When building our FNN, we first reserve $20\%$ of the error-free data for model testing. Of the remaining data, three quarters are randomly chosen for training, and the rest are used for validation. We use the validation-set approach (see \cite[\S 5.1]{Hastie2017}) to select values of the hyper-parameters, such as the number of hidden layers and nodes, regularization coefficients $\alpha_w$ (for weights) and $\alpha_b$ (for bias terms), and activation functions.

The candidate hidden layer structures are $\{$(112), (256), (140,112), (140,112,84)$\}$, the candidate activation functions are $\{$sigmoid, tanh$\}$, the candidates of both the bias regularization and weight regularization hyper-parameters are $\{$0.01, 0.001$\}$, and the candidate norms for error terms in the loss are $\{L1, L2\}$. So there are $4\times2\times2\times2\times2=64$ different sets of hyper-parameter values. For each hidden layer structure (hidden layer number and node number in each layer) we report the two models with top validation performances in Table \ref{table1}. To make sure the model outputs lie in the same range of the dependent variables, which is $[0,100]$, we add a sigmoid activation (whose output lies in $(0,1)$) to each node in the output layer, and then multiply the node output by a factor of 100.

\begin{table}[!htb]
\caption{Model performances with different combinations of hyper-parameters and activations. }
\label{table1}
\centering%
\begin{tabular}{ p{3cm}  p{2cm}  p{2cm} p{1cm} p{1cm}p{2cm}p{2cm}  }
\toprule
Hidden Layers and Nodes &Loss &Activation  & $\alpha_b$ & $\alpha_w$ &Train $R^2$ &Validat $R^2$\\
\toprule
$(112)$            	&MSE	     &sigmoid       &0.001   &0.001   &$94.4\%$        &$92.7\%$ \\
$(112)$              &MSE		&sigmoid       &0.01   &0.001   &$94.7\%$         &$92.8\%$ \\
$(256) $             &MSE		&sigmoid      &0.01    &0.001   &$97.8\%$         &$96.0\%$ \\
$(256) $             &MSE		&tanh           &0.01   &0.001   &$97.9\%$         &$96.1\%$ \\
$(140,112) $      &MSE		&tanh           &0.01      &0.001   &$98.4\%$         &$96.6\%$ \\
$(140,112) $      &MSE		&sigmoid           &0.001    &0.001    &$98.5\%$         &$96.7\%$ \\
$(140,112,84)$  &MSE		&tanh           &0.001      &0.001    &$98.5\%$         &$96.8\%$ \\
$(140,112,84)$  &MSE		&sigmoid           &0.01    &0.001    &$98.8\%$         &$97.0\%$ \\
\bottomrule
\end{tabular}
\end{table}

\begin{remark}
The hidden layers structure is expressed with a tuple, each entry represents the number of nodes in a hidden layer. For example, (84,56) in Table \ref{table1} indicates that the model has two hidden layers, where the first one has 84 nodes and the second one has 56 nodes. Except the ones in the input layer, each node is associated with a bias node. The regularization coefficient for weights is $\alpha_w$ and the regularization coefficient for the bias term is $\alpha_b$. ``Train $R^2$'' represents the $R^2$ of model predictions on the training data, and ``Validat $R^2$'' represents the $R^2$ of model prediction on the validation data.
\end{remark}

In Table \ref{table1}, we see that the winner of validation performances is the model in the last row, which has a hidden layer structure of (140,112,84). Its activation function is sigmoid, the error norm in its loss function is $L2$, $\alpha_b=0.01$ and $\alpha_w=0.001$.

Note that one drawback of the validation-set approach comparing to cross-validation (an alternative method for selecting hyper-parameter values, see, e.g. \cite[\S 7.10]{Hastie2009} ) is that its validation performance might be unstable -- it may change when the training and validation sets are reassigned. To see whether this drawback is true for our winner hyper-parameter value set, we perform cross-validation on it. More specifically, we do the following.

\begin{enumerate}
\item Combine the traning and validation samples to one set $\mathcal{D}$.
\item Randomly divide $\mathcal{D}$ to five parts, $\{\mathcal{D}_1, \mathcal{D}_2,...,\mathcal{D}_5 \}$, with equal sizes.
\item For each $i\in{1,2,...,5}$, train the model with $\cup_{j\in\{1,2,...,5\}, j\neq i}\mathcal{D}_j$ and validate with $\mathcal{D}_i$. The validation $R^2$ is denoted as $R^2_i$.
\item Take the average of $\{R^2_i\}_{i=1}^5$ and denote it as $\bar{R}^2$.
\end{enumerate}
The results of the above process are reported in Table \ref{table2}. We can see that $\bar{R}^2$ is $96.4\%$, which is close to the previous validation $R^2$ of $97.0\%$ in Table \ref{table1}. Therefore, the winner hyper-parameter value set does not suffer from the drawback under concern.

\begin{table}[H]
\caption{Cross validation of the winner hyper-parameter value set in Table \ref{table1}. Procedures 1-4 in \ref{buildmodel} are performed.}
\label{table2}
\centering%
\begin{tabular}{ p{3cm}  p{3cm}p{3cm}  }
\toprule
Fold number                 & Train $R^2$  & Validation $R^2$\\
\toprule
Fold 1      				 &$98.0\%$        &$96.3\%$ \\
Fold 2      				 &$98.4\%$        &$96.7\%$ \\
Fold 3      				 &$98.2\%$        &$96.0\%$ \\
Fold 4       			&$98.5\%$        &$96.6\%$ \\
Fold 5      				 &$98.1\%$        &$96.5\%$ \\
Average ($\bar{R}^2$)   			&$98.2\%$        &$96.4\%$\\
\bottomrule
\end{tabular}
\end{table}

\begin{figure}[!htb]
	\centering
		\includegraphics[scale=.30]{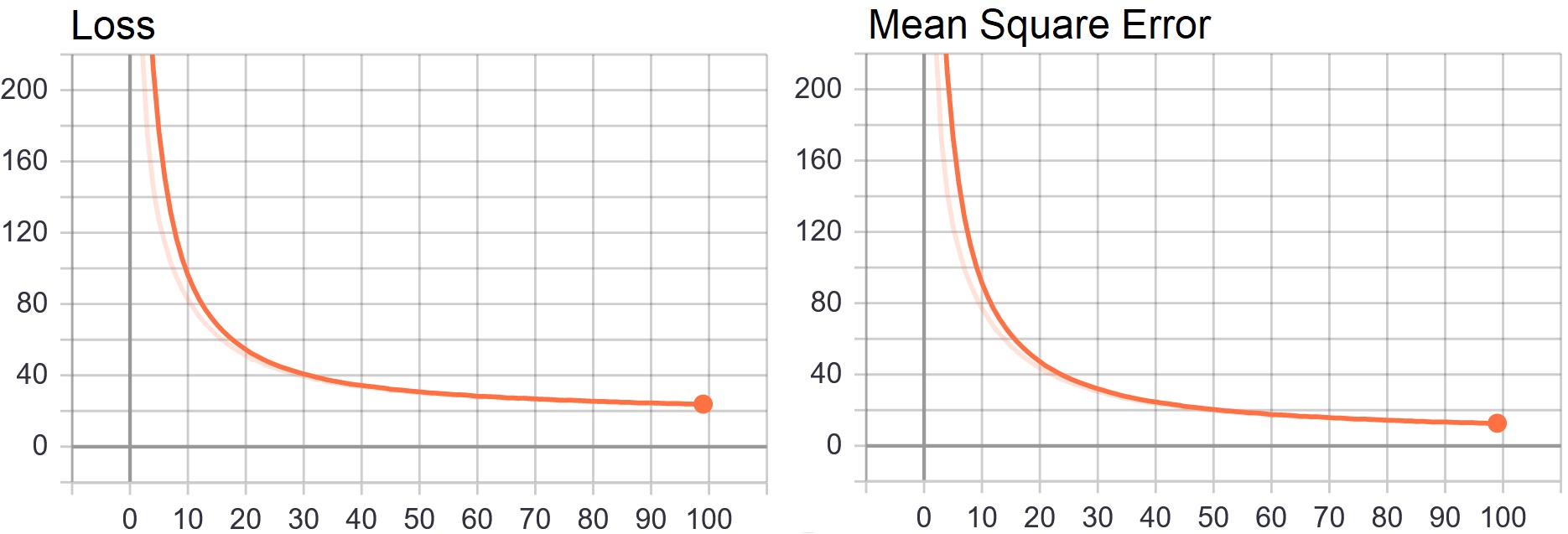}
\caption{The loss and MSE in the training process. }
\label{Fig:loss}
\end{figure}

\begin{figure}[!htb]
	\centering
		\includegraphics[scale=.35]{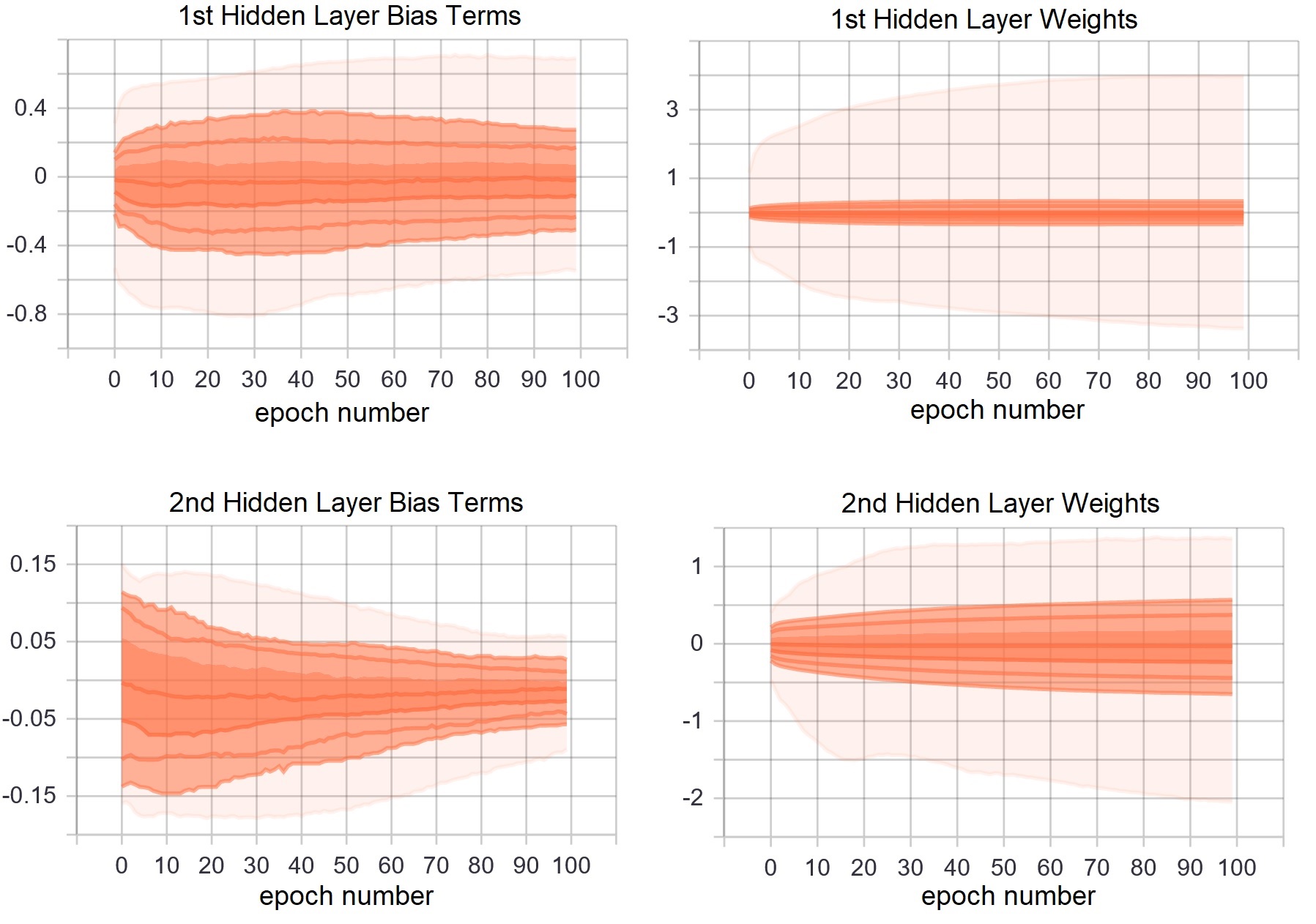}
		\includegraphics[scale=.35]{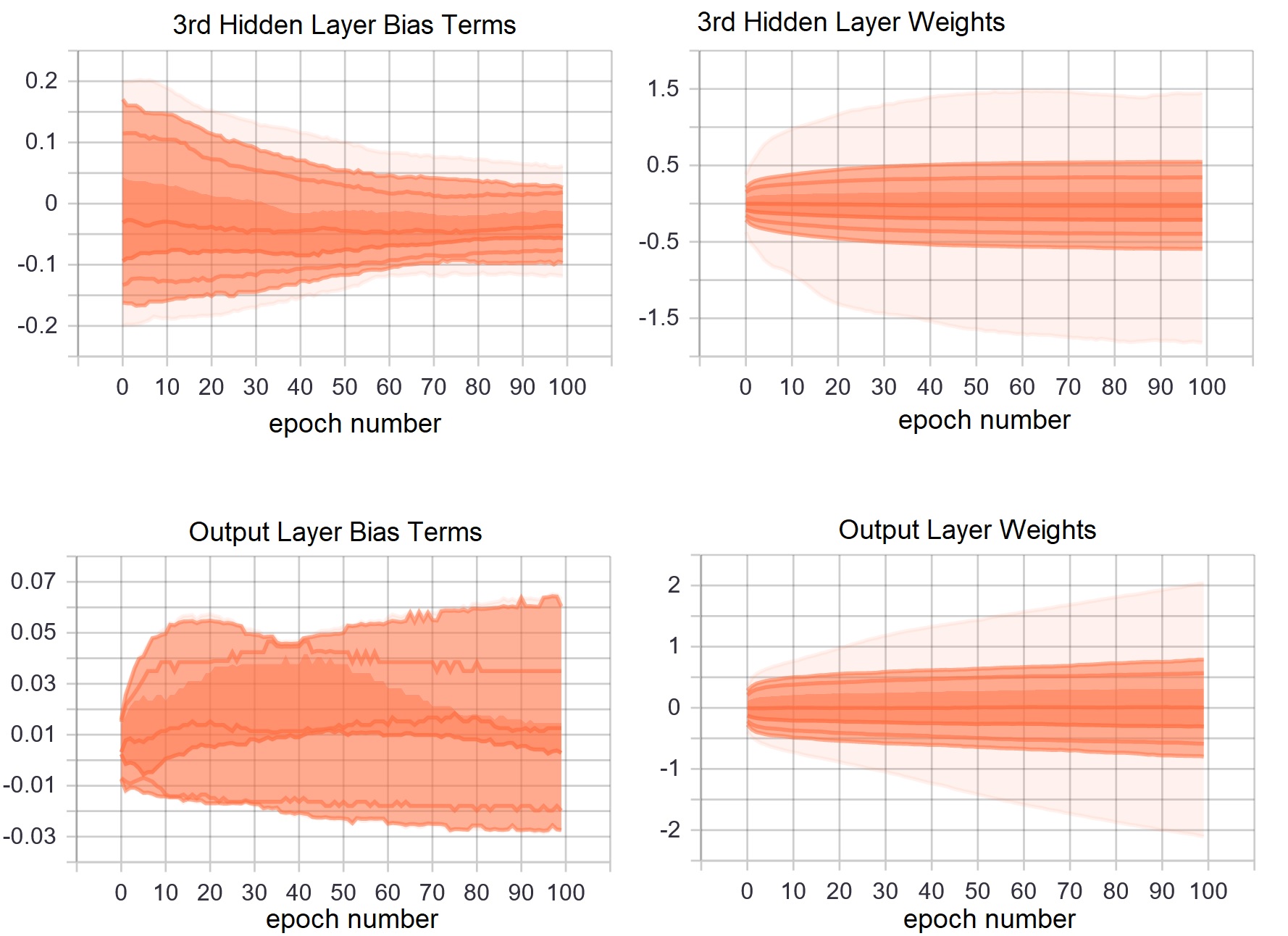}
\caption{The distribution of bias terms and weights in the training process.}
\label{Fig:weight}
\end{figure}

Now we re-train the model with the winner hyper-parameter value set with $\mathcal{D}$ and compute its performance on the test set. The training and test $R^2$s are $98.18\%$ and $98.33\%$, respectively. Since these two $R^2$'s are almost equal to each other, we believe there is no overfitting problem. The training process is shown in Figures \ref{Fig:loss} and \ref{Fig:weight} through TensorBoard\footnote{TensorBoard is a tool from TensorFlow to plot statistics for neural net training process.} plot. The left graph in Figure \ref{Fig:loss} shows that by the gradient descent algorithm the loss is decreasing as the number of epochs increase, where the bias terms and weights are updated once in each epoch. The right graph shows how the error terms\footnote{It is the loss without the regularization part} decreases. In Figure \ref{Fig:weight}, the eight plots show the distribution of the bias terms and weights in each epoch. For example, the last plot in this figure shows that the distribution of output layer weights starts from about $[-0.4, 0.4]$, and as the epoch number increases the distribution interval also enlarges gradually to approximately $[-0.8, 0.8]$.

To see the model's prediction accuracy for each of the eight dependent variables, we report the testing $R^2$ for each of them in Table \ref{table3}.

\begin{table}[H]
\caption{Testing $R^2$ for each of the 8 entries in the output.}
\label{table3}
\centering%
\begin{tabular}{ p{5cm}  p{5cm}  }
\toprule
The $i^{th}$ entry & $R^2$  \\
\toprule
1st entry & $98.56\%$\\
2nd entry & $98.00\%$\\
3rd entry & $98.72\%$\\
4th entry & $98.27\%$\\
5th entry & $98.63\%$\\
6th entry & $98.42\%$\\
7th entry & $98.69\%$\\
8th entry & $97.38\%$\\

\bottomrule
\end{tabular}
\end{table}

\subsubsection{Robustness of the model for data with errors}
\label{noise-method}

Measurement errors are common in scientific experiments. So when applied in practice, the inputs to our model are likely to contain errors. To show the robustness of our model in the cases of input data errors, we report the model performance on each of the following four scenarios of noised testing data.
\begin{enumerate}
\item \emph{Normally-distributed errors}. For each testing sample $(\mathbf{x},\mathbf{y})$, we add noises to the first 800 entries of $\mathbf{x}$. More specifically, suppose $\mathbf{x}=(x_1, x_2, ..., x_{802})$. We independently sample $\epsilon$ from $N(\mu,\sigma^2)$ 800 times and obtain the error vector $\varepsilon=(\epsilon_1, \epsilon_2,...,\epsilon_{800})$, then we replace $x_i$ with $x_{i}(1+\epsilon_i)$ for $i=1,2,...,800$. Note that normalization will still be performed to the noised samples before feeding them to the model.

\item \emph{Uniformly-distributed errors}. In this scenario, the testing data are modified in the same way as the normally-distributed-error scenario, except that the normal distribution $N(\mu,\sigma^2)$ is replaced with the uniform distribution.

\item \emph{Poisson errors}. In this scenario, the testing data are modified in the same way as the normally-distributed-error scenario, except that the normal distribution is replaced with a Poisson-distributed noise divided by 100.

\item \emph{Time lag}. In this scenario, there is a time lag/advancement between the measured input and the true value. For each testing data $\mathbf{x}$, we randomly sample (with replacements) the time lag $\tau$ from $\{-m,-m+1,...,m-1,m\}$, where $m$ is a positive integer and each number has the same probability to be chosen. Then we shift $\mathbf{x}$ in time by $\tau$, the missing entries will be replaced with zero. Note that the last two entries of $\mathbf{x}$ will not be shifted.
\end{enumerate}

The performance of our model (the one in Table \ref{table3}) on the testing samples in each of the four scenarios are reported in Table \ref{table4}. The results show that the model performs poorly on the data shifted in time. To fix this problem, we re-build the model in the next section from training and validation samples that are shifted in time.

\begin{table}[H]
\caption{$R^2$ on data with errors.}
\label{table4}
\centering%
\begin{tabular}{ p{5cm}  p{3cm}  }
\toprule
Error type & $R^2$  \\
\toprule
N$(0.04, 0.1^2)$. & $94.86\%$\\
Uniform$(-0.2, 0.1)$. & $93.35\%$\\
Poisson$(5)$/100. & $95.69\%$\\
Max-shift is 1. & $47.94\%$\\
\bottomrule
\end{tabular}
\end{table}

\subsection{Building the Model VIP-FNN on Shifted Artificial Data and Real Data}
\label{noisedata}

From Table \ref{table4} we see that our previous FNN performs poorly when data shifts in time, which often happens in laboratory experiments due to measurement errors. Moreover, our previous training samples do not contain real data obtained from laboratory experiments, which may contain noises with unknown structures. So to increase the model's capacity to predict with data in the real world, we train and validate a new model on simulated data shifted in time and real data. This solution model will be called as the VIP-FNN.

Specifically, the new dataset $\mathcal{D}$ consists of two parts, $\mathcal{D}_1$ and $\mathcal{D}_2$. The first part $\mathcal{D}_1$ consists of all the 63500 data samples in Section \ref{data_structure}, but each $\mathbf{x}$ (the first 800 entries) is shifted with a random time-lag uniformly distributed in $\{-8,-7,$ $...,7,8 \}$. The second part $\mathcal{D}_2$ are constructed with five injection informed data $\{\mathbf{x}^{(r)},r=1,2,...,5\}$ from the real world experiments. A detailed description of the real-world data can be found in Section \ref{Realdata}. For each $\mathbf{x}^{(r)}$, we use the traditional regularization method to obtain $\mathbf{y}^{(r)}$, the vector of values of the eight parameters
\begin{equation*}\label{parameterTradition}
\mathbf{y}^{(r)}=[a_{\Rmnum{1},1}, b_{\Rmnum{1},1}, a_{\Rmnum{2},1}, b_{\Rmnum{2},1}; a_{\Rmnum{1},2}, b_{\Rmnum{1},2}, a_{\Rmnum{2},2}, b_{\Rmnum{2},2}] = [9.54, 0.91, 9.53, 1.00;
2.74, 0.43, 1.80, 0.08].
\end{equation*}
Note that $\mathbf{y}^{(r)}$ are the same for all $r=1,2,...,5$, which corresponds the intrinsic quantity of adsorption isotherms for a fixed chemical system. The pair $(\mathbf{x}^{(r)},\mathbf{y}^{(r)})$ is included in $\mathcal{D}_2$. So there are in total five samples in $\mathcal{D}_2$. Then, the training, validation, and testing datasets are constructed in the following steps.

\begin{enumerate}
\item \textit{Testing Samples.} $20\%$ of $\mathcal{D}_1$ constitutes the testing samples and are removed them from $\mathcal{D}_1$.

\item \textit{Training Samples.} $75\%$ of the remaining samples in $\mathcal{D}_1$ are randomly selected, and three samples from $\mathcal{D}_2$ are also included in the training samples. In order to increase the weights of real data, for the training samples we add duplicates of the three samples from $\mathcal{D}_2$ so that their portion to training samples from $\mathcal{D}_1$ is $1:10$.

\item \textit{Validation Samples.} The remaining samples in $\mathcal{D}_1$ are included in the validation sample set (they constitute $80\%\times25\%=20\%$ of the original $\mathcal{D}_1$). The two samples left in $\mathcal{D}_2$ are included in the validation set. Again, to increase their weights, duplicates of these two samples are added so that their portion to validation samples from $\mathcal{D}_1$ is $1:10$.
\end{enumerate}

We follow the same procedure in Section \ref{buildmodel} to select hyper-parameter values and activation functions. The hidden layer structures under consideration are $$\{ (140,112,84), (140,112,112,112), (168,168,168,168,112),(168,168,168,168,168,112)  \}.$$
The candidate values for other hyper-parameters remain. The performances of these candidate models on the training data are reported in Table \ref{table5}. The winner model is the one with hidden layers (168,168,168,168,168,112) with a training $R^2$ of $96.0\%$ and a validation $R^2$ of $93.3\%$.

\begin{table}[!htb]
\caption{Performances of candidate models on time-shifted data. For each hidden layer structure, we report two candidate models with top validation performances,  The regularization coefficient for weights is $\alpha_w$ and the regularization coefficient for the bias term is $\alpha_b$, both take values from $\{0.01, 0.001\}$. The activation functions under considerations are tanh and sigmoid. The candidate norms in the loss are $L1$ and $L2$. Train $R^2$ represents the $R^2$ of model predictions on the training data, and Validat $R^2$ represents the $R^2$ of model prediction on the validation data.}
\label{table5}
\centering%
\begin{tabular}{ p{5cm} p{1cm} p{2cm} p{0.75cm} p{0.75cm} p{1.8cm} p{1.9cm}  }
\toprule
Hidden Layers and Nodes &Loss &Activation & $\alpha_b$ & $\alpha_w$ &Train $R^2$ &Validat $R^2$\\
\toprule

$(140,112,84)$                  &L2           &sigmoid       &0.01   &0.001   &$93.1\%$         &$86.4\%$ \\
$(140,112,84)$                  &L2		     &sigmoid       &0.001    &0.001   &$92.9\%$         &$86.5\%$ \\
$(140,112,112,112)$          &L2 		&sigmoid      &0.001    &0.001   &$95.1\%$        &$88.3\%$ \\
$(140,112,112,112)$          &L2 		&sigmoid      &0.01    &0.001   &$94.6\%$          &$90.1\%$ \\
$(168,168,168,168,112)$         &L2		     &sigmoid           &0.01     &0.001   &$96.3\%$         &$90.1\%$ \\
$(168,168,168,168,112)$         &L2		     &sigmoid      &0.001    &0.001    &$97.0\%$      &$92.9\%$ \\
$(168,168,168,168,168,112)$  &L2			&sigmoid           &0.01      &0.001    &$95.2\%$         &$91.9\%$ \\
$(168,168,168,168,168,112)$  &L2			&sigmoid           &0.001      &0.001    &$96.0\%$         &$93.3\%$ \\
\bottomrule
\end{tabular}
\end{table}

Table \ref{table6} reports the cross-validation performance of the winner hyper-parameter value set in Table \ref{table5}. The validation $R^2$'s of these models are all equal to or above $87.8\%$, and the average is $92.4\%$, which is close to the validation $R^2$ of $93.3\%$ in Table \ref{table5}. This implies that the performances of models with the winner hyper-parameter value set in Table \ref{table5} are stable and reliable. So in the following we stick to this hyper-parameter value set, and train a new model on the combined training and validation sets. The training $R^2$ and testing $R^2$ of this new model are $96.42\%$ and $93.97\%$, respectively. Table \ref{table7} reports the prediction accuracies for each of the 8 entries on the testing data.

\begin{table}[H]
\caption{Cross validation of the winner model in Table \ref{table5}. Procedures 1-4 in \ref{buildmodel} are performed on the combined training and validation datasets constructed at the beginning of Section \ref{noisedata}. We make sure that for each fold, three real samples (randomly selected) and their duplicates are included in the training set. The two remaining real samples in $\mathcal{D}_2$ and their duplicates are included in the validation set. So there are no overlaps between the training and validation datasets for each fold. The portion of simulated samples to real samples (and duplicates) is 10:1. }
\label{table6}
\centering%
\begin{tabular}{ p{3cm}  p{3cm}p{3cm}  }
\toprule
Fold number                 & Train $R^2$  & Validation $R^2$\\
\toprule
Fold 1      				 &$96.2\%$        &$93.5\%$ \\
Fold 2      				 &$97.0\%$        &$94.1\%$ \\
Fold 3      				 &$96.7\%$        &$94.3\%$ \\
Fold 4       			&$91.6\%$        &$87.8\%$ \\
Fold 5      				 &$96.2\%$        &$92.5\%$ \\
Average ($\bar{R}^2$)   			&$95.5\%$        &$92.4\%$\\
\bottomrule
\end{tabular}
\end{table}

\begin{table}[!htb]
\caption{Testing $R^2$ for each of the 8 entries in the output. This table is for the model with the winner hyper-parameter value set in Table \ref{table5}, which is trained with the training and validation datasets. These two data sets and the testing set are constructed at the beginning of Section \ref{noisedata}.}
\label{table7}
\centering%
\begin{tabular}{ p{5cm}  p{5cm}  }
\toprule
The $i^{th}$ entry & $R^2$  \\
\toprule
1st entry & $95.37\%$\\
2nd entry & $92.33\%$\\
3rd entry & $94.82\%$\\
4th entry & $92.83\%$\\
5th entry & $95.45\%$\\
6th entry & $92.98\%$\\
7th entry & $95.11\%$\\
8th entry & $92.93\%$\\
\bottomrule
\end{tabular}
\end{table}

At the end of this subsection, let us show the robustness of our new model on data with more noises. To this end, we add simulated measurement errors to the new testing data (constructed at the beginning of \ref{noisedata}) in the same way as in \ref{noise-method}. The performances of the new model\footnote{Note that the model in Table \ref{table7} is not re-trained for Table \ref{table8}.} in Table \ref{table7} on these noised data are included in Table \ref{table8}. All of the $R^2$s are not far from the $R^2$ on the noise-free testing samples (time-shifted), implying that the model is robust against normal, uniform, and Poisson noises in data.

\begin{table}[!htb]
\caption{$R^2$ on data with noises of different type. Similar to Table \ref{table4}, we add simulated measurement errors to the new testing data and show the performance of our new  model on these noised data.}
\label{table8}
\centering%
\begin{tabular}{ p{5cm}  p{3cm}  }
\toprule
Error type & $R^2$  \\
\toprule
Error-free &$93.97\%$ \\
N$(0.04, 0.1^2)$. &$83.90\%$ \\
Uniform$(-0.2, 0.1)$. & $82.67\%$\\
Poisson$(5)/100$. & $81.39\%$\\
\bottomrule
\end{tabular}
\end{table}

\section{Real Data Application}
\label{Realdata}

Now, we are ready to employ our developed VIP-FNN to the real world problem. All the injection informed data from real world experiments, including the ones from which $\mathcal{D}_2$ is constructed, can be found in the simulation section in \cite{zhang2016regularization}. For readers' convenience, we briefly describe the experimental information as follows: two pharmaceutical substances (propranolol and alprenolol), were found to separate well on a Kromasil C18 column at 25 $^{\circ}$C using a mobile phase composed of 28:72 (v/v) acetonitrile: aqueous phosphate buffer (pH 2.54, ion strength 0.1). Seven binary (propranolol and alprenolol) elution profiles with $g(x)\equiv [0, 0]^T \textrm{mM}$ and $h^\sigma(t)$ equal to [5, 5], [0.75, 0.75], , [10, 10], [15, 15], [15, 5], [5, 15], and [30, 30]mM, respectively, were recorded. Other parameters were as follows: $L=15$ cm (inner diameter, 0.46 cm), $F = 0.78$, $u = 0.125$ cm/s, $N_x = 9000$ and injection volume $= 50\mu l$. The elution profiles were recorded using an Agilent 1100 Chemstation LC instrument with a UV detector.

\begin{figure}[!htb]
	\centering
		\includegraphics[scale=.46]{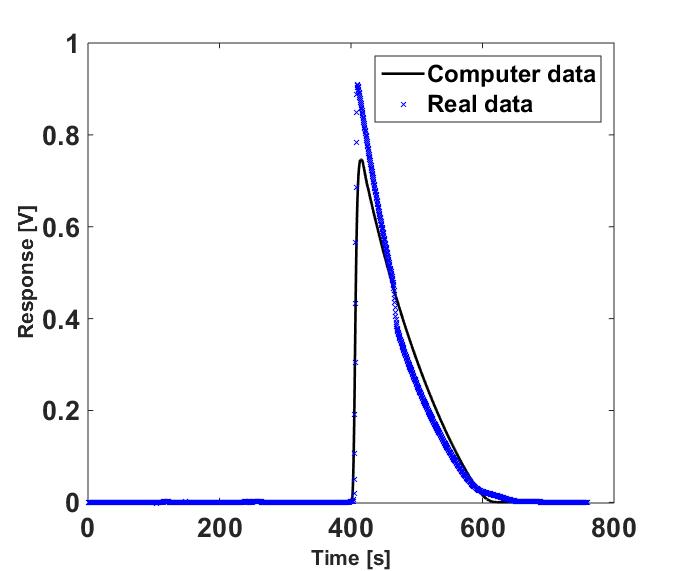}
		\includegraphics[scale=.46]{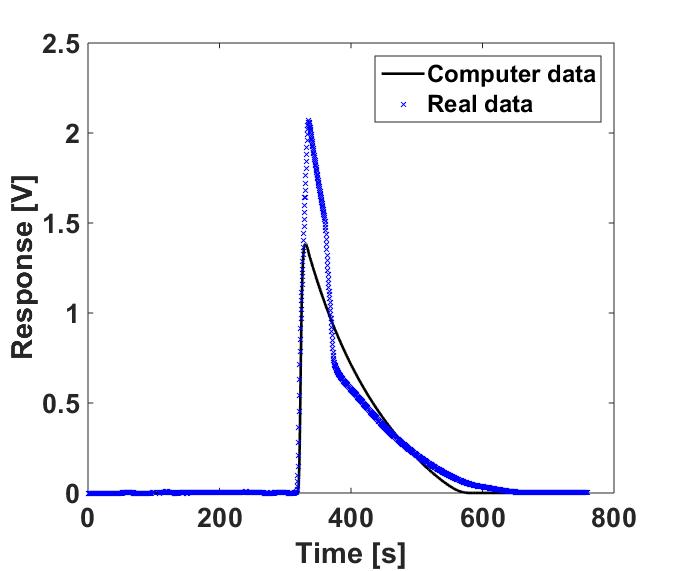}
\caption{Tests with real data.}
\label{Fig:RealData}
\end{figure}

Five groups of data have been used as the training or validation data in the building of our VIP-FNN in Section \ref{noisedata}. The rest two groups of data, corresponding to the injection profiles [5, 15] and [30, 30]mM, are used in here for testing our method. The experimental elution profiles corresponding to these two injections of propranolol and alprenolol are displayed in Fig.~\ref{Fig:RealData} (``$\times$''). The estimated adsorption isotherm parameters for two groups of data by VIP-FNN are
\begin{equation*}
\hat{\mathbf{y}}_1=[a_{\Rmnum{1},1}, b_{\Rmnum{1},1}, a_{\Rmnum{2},1}, b_{\Rmnum{2},1}; a_{\Rmnum{1},2}, b_{\Rmnum{1},2}, a_{\Rmnum{2},2}, b_{\Rmnum{2},2}] = [10.90, 1.08, 9.94, 1.77, 3.53, 1.03, 2.27, 0.12]
\end{equation*}
and
\begin{equation*}
\hat{\mathbf{y}}_2=[a_{\Rmnum{1},1}, b_{\Rmnum{1},1}, a_{\Rmnum{2},1}, b_{\Rmnum{2},1}; a_{\Rmnum{1},2}, b_{\Rmnum{1},2}, a_{\Rmnum{2},2}, b_{\Rmnum{2},2}] = [9.31, 0.81, 10.50, 0.68, 2.29, 0.29, 2.11, 0.12],
\end{equation*}
respectively. As we can see, these two groups of estimators are very close to each other. This result verifies that the two different types of datasets correspond to the same adsorption isotherm, which is the intrinsic property of two substances (propranolol and alprenolol). The difference between these two estimators is mainly caused by the noise. In practice, we can use the average of the these two estimators as the final estimated adsorption isotherm parameters
\begin{equation*}\label{parameterReal}
\hat{\mathbf{y}}=[a_{\Rmnum{1},1}, b_{\Rmnum{1},1}, a_{\Rmnum{2},1}, b_{\Rmnum{2},1}; a_{\Rmnum{1},2}, b_{\Rmnum{1},2}, a_{\Rmnum{2},2}, b_{\Rmnum{2},2}] = [10.10, 0.94, 10.22, 1.23; 2.91, 0.66, 2.19, 0.12].
\end{equation*}
The solid curve in Fig.~\ref{Fig:RealData} are simulated response signal at the outlet with the estimated parameters $\hat{\mathbf{y}}$. We can see that they are close to the real data (``$\times$''). implying that the implementation of our VIP-FNN to this real-world problem is successful.

At the end of this section, we point out that the developed VIP-FNN method has a number of advantages over the traditional methods. For example, its implementation is simple and fast. To obtain the interested adsorption isotherm parameter values, we only need to input the measured data and injection profile to the model. The results will be output in seconds,  which usually takes at least half an hour with traditional regularization algorithms. Moreover, the VIP-FNN method is robust against the issues with traditional variational regularization methods, whose results strongly depend on initial guesses of adsorption isotherm parameters and the choice of regularization parameters. Also, our VIP-FNN is "experienced" -- it is trained with a large number of samples from both the real world and the recognized physics model \eqref{eq:prime} with different injection profiles, which helps the model to accurately estimate the adsorption isotherm parameters from new data with only few injections.

\section{Conclusions}
\label{sec:Con}

In this work, we have proposed a novel data-driven knowledge-aided framework, named as Virtual Injection Promoting Feed-forward Neural Network (VIP-FNN), for estimating adsorption isotherm parameters in liquid chromatography. The numerical results with both synthetic and experimental data show that it is competitive
with state-of-the-art benchmarks. Although more experience with other systems is needed to fully understand the potential and the limitations of VIP-FNN, the initial applications to model problems and the real propranolol and alprenolol system are promising, yielding excellent fits of the data with minimum time cost. We therefore believe VIP-FNN will be a useful tool to study the adsorption isotherms of chromatography. Moreover, the methodology used in this paper can also be applied to solve other inverse problems in partial differential equations with sparse boundary data.

\section{Acknowledgement}


This work of Y. Zhang is supported by the Guangdong fundamental and Applied Research Fund [No. 2019A1515110971] and the Swedish Knowledge Foundation (KKS) project Synergy [No. 20170059].

\section*{Appendix: Structure of the Feed-forward Neural Network}
\label{Appendix}

This appendix introduces what is Feed-forward Neural Network (FNN) used for, its model structure, and how the
model is trained validate.

\begin{figure}[!t]
	\centering
		\includegraphics[scale=.5]{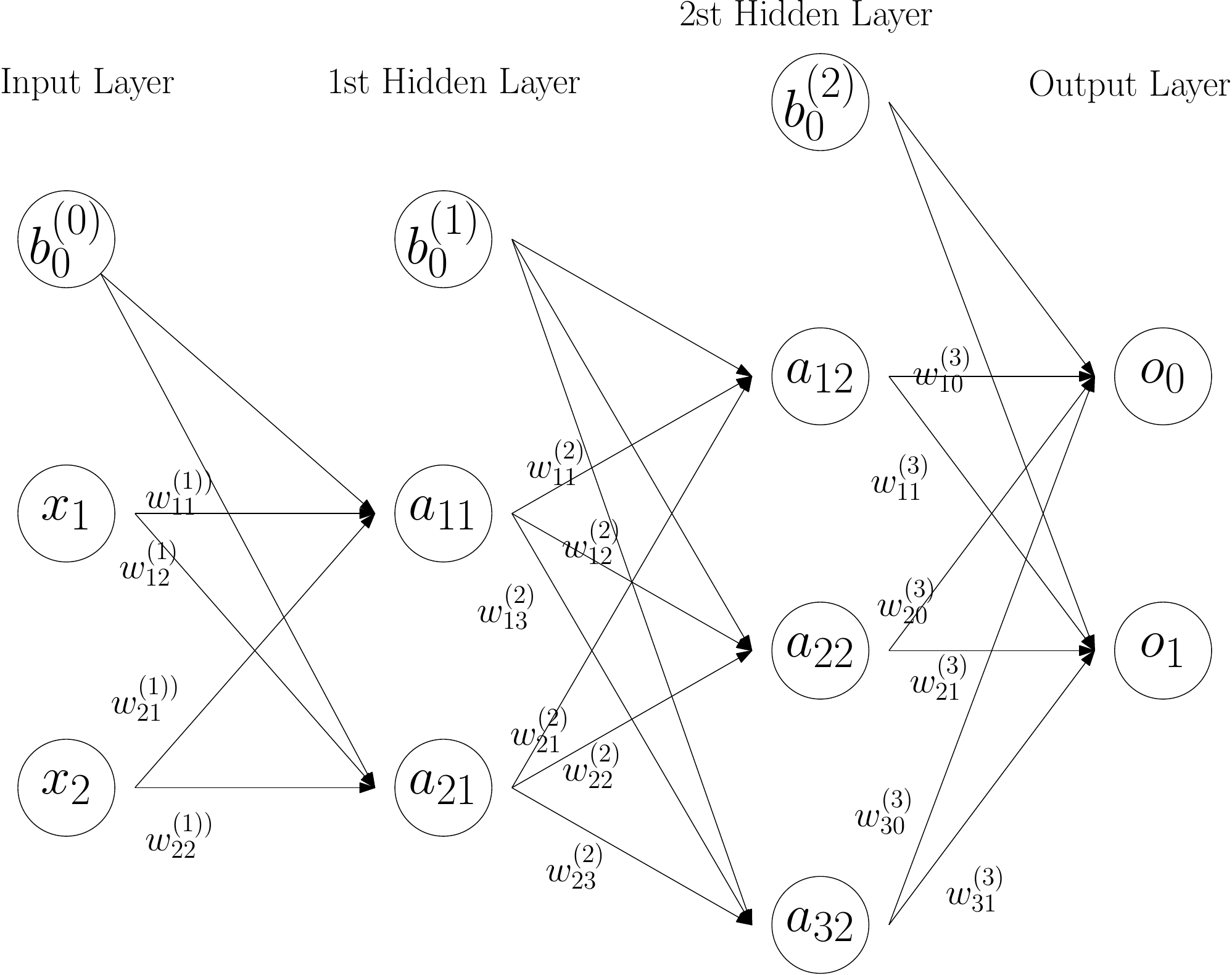}\\
	\caption{An FNN Example. The feed-forward neural network in this example has two hidden layers. The first hidden layer has two nodes and the third one has three nodes. The superscript denotes the hidden layer number.}
	\label{FNNexample}
\end{figure}

An FNN contains an input layer that takes in the independent variables (also called features) values, an output layer that produces values for model predictions, and the hidden layers (if any) whose structure is controlled by hyper-parameters set by users. Figure \ref{FNNexample} is an example of FNN for a regression problem with two independent variables $\{X_0, X_1\}$ and two dependent variables $\{Y_0, Y_1\}$, where
\begin{itemize}
\item the value pair $\{x_0, x_1\}$ is a realization of $\{X_0, X_1\}$ and also the input to the model;
\item The $b$ terms (called bias terms) and the weights $\{w_{ij}\}$ are parameters. The node values (except the bias nodes and the ones in the input layer) are transformations of the weighted sums of the outputs from the previous layer, i.e.
\[ a_{ij} :=   g\left(w_{0i}b_0^{(j-1)}+\sum_{k=1}^{n_{j-1}}w_{ki}^{(k)}a_{k,j-1} \right),\quad j=1,2,3, \]
where $g$ is called the \textit{activation function} which is pre-selected, $n_{j-1}$ denotes the number of nodes in the $(j-1)^{th}$ layer. $a_{ij}=x_i$ if $j=0$, and $a_{ij}=o_i$ if $j=3$. Normally, the activation is the same for every node except the ones in the output layer. For classification problems, the activation function for the output layer nodes may be the softmax function, which converts real numbers to probabilities of classes. While for regression problems, the activation function for the output layer may just be an identity function.
\item The $o$ terms are outputs from the neural net. They are supposed to match the data for the dependent variables.
\item Note that this example has only one bias term in each layer (except the input layer). Other neural networks may have one bias associated with each node.
\end{itemize}
The role of the activation function is to control the amount of contribution made by the corresponding node to the model's output, and there are different choices for this function. The widely used ones are sigmoid, ReLU, and tanh, each having its own advantages and disadvantages. For example, ReLU is computationally efficient to use and can avoid vanishing gradient but it may lead to the problem of ``dead neuron", see Remark \ref{remarkReLu} for details.

To find values of the parameters (the weights and bias terms) so that the model's outputs are close to data of the dependent variables, a process called \textit{model training} is performed in the following procedures.
\begin{enumerate}
\item Pre-select hyper-parameters such as the loss function and the hidden layer structure.
\item Assign initial values to the parameters (weights and bias terms).
\item Compute the patial derivative of the loss with respect to each weight and bias term through the back-propagation algorithm.
\item Based on the partial derivatives, update the parameter values to decrease the loss value.
\item Repeat step 3-4 until the loss converges (means it does not decrease any more) or reach some threshold.
\end{enumerate}
The number of hidden layers and nodes are treated as hyper-parameters which can be adjusted according to the model's performance on the validation data. Refer to Chapter 11 of \cite{Hastie2009} for more details on neural network.

\end{document}